\documentstyle{article}

\begin{document}

\newtheorem{theorem}{\bf Theorem}
\newtheorem{proposition}{\bf Proposition}
\newtheorem{remark}{\bf Remark}
\newtheorem{corollary}{\bf Corollary}
\newtheorem{definition}{\bf Definition}
\newtheorem{lemma}{\bf Lemma}
\newtheorem{claim}{\bf Claim}
\newtheorem{example}{\bf Example}
\newtheorem{hypothesis}{\bf Hypothesis}

\title
{\textbf{On the first determination of Mercury's perihelion advance}}
\author{Diana R. CONSTANTIN}
\date{February 25, 2010}
\maketitle

\begin{abstract}
The first determination of the perihelion advance of Mercury's orbit was obtained by Leverrier from the analysis of the transit contacts of the planet on the solar disk. He obtained for the advance the value $\delta\pi'=38".3/century$, considering that the value $\delta e'$, namely the correction of the variation of the planet's orbit eccentricity, is negligible. In this paper $\delta\pi'$ and $\delta e'$ are calculated by the least squares method, on the basis of the meridian observations used by Leverrier. Thus, we obtain for advance the value $\delta\pi'= 42".8/century$, which is close to the one given in the theory of general relativity. The same, we obtained the value $\delta e'= -0".044/year$, which is lower in absolute value than Leverrier's estimation $\delta e'= -0".0806/year$.
\end{abstract}



\section{Introduction}\label{sec1}

The first determination of Mercury's advance was obtained by Leverrier and it was published in 1859 in $\it{Annales  \,de\, l'Observatoire\, de\, Paris}$ \cite{5}. To this end he used the observations of the planet's transits over the solar disk of the period 1661-1848. Mercury's transits over the solar disk take place in November, in the vicinity of  the ascending node of the planet's orbit, and in May, in the vicinity of the descending node.

In the vicinity of the nodes the true heliocentric longitude $\it{v}$ is practically reduced to the sum of the mean longitude and of the center's equation, and has the following form:
$\it{v = f(l,\pi,e)= f(\varepsilon +nt,\pi,e)}$,
where the mean longitude is $l=\varepsilon +nt$, $\varepsilon$ being the mean longitude at the epoch and $\it{n}$ the mean motion \cite{2}. Hence the correction of $\it{v}$ is of the form
\vspace{0.3cm}

$\hspace{1.5cm}\it{\delta v = \frac{\partial f}{\partial l}(\delta\varepsilon +t \delta n)+
\frac{\partial f}{\partial \pi}(\delta \pi +t \delta \frac{d\pi}{dt})+\frac{\partial f}
{\partial e}(\delta e +t \delta \frac{d e}{dt})}.$
\vspace{0.2cm}

For the 8 transits in November Leverrier obtained small corrections of the longitude $\it{v}$ of Mercury, while  for the 5 transits of May  he obtained much more important corrections, which actually increase in time \cite{2},\cite{5}.
Thus, he made a first determination of the advance by considering a linear time dependence for $\delta \it{v}$, namely: for the November transits $\delta \it{v} = a + tb$ and for May transits $\delta \it{v} = a' + tb'$. He obtained the relation:
$\delta\pi' + 2.72\,\delta e'= 0".392$,
where $e'=\frac{d e}{dt}$ and $\pi'=\frac{d \pi}{dt}$.
After a more precise calculation he obtained:
\begin{equation}\label{1.4}
\delta\pi' + 2.72\,\delta e'= 0".383/year.
\end{equation}

Leverrier considered the value of $38".3/century$ as perihelion advance, neglecting $\delta e'$ \cite{5}. He motivates this by resorting to the 397 meridian observations on Mercury made at Paris Observatory in the periods 1801-1828 and 1836 -1842. By processing of these approx. 400 meridian observations, he obtained 195 equations of condition. Analyzing a part of the 195 equations of condition corresponding to the planet's apparent geocentric longitude, he obtained $\delta e'= -0".0806$ and on the basis of relation (\ref{1.4}), reached the conclusion that $\delta\pi'= 60"/century$.

Forward, the Leverrier's result $\delta\pi'= 38".3/century$ was also confirmed and specified by Newcomb, who obtained $\delta\pi'= 41".24/century $ \cite{6} by using in his determinations both the planet's transit over the solar disk and meridian observations. The problem was also studied by Clemence, whose analysis led to an advance of $\delta\pi'= 43".03/century $ \cite{3},\cite{4}.

Coming back to the Leverrier's work described above, namely to this deviation of $60"/century $ as against the value of $38".3/century $, we have tried to determine the value of $\delta e'$ more precisely,  by solving the system of the 195 equations of condition obtained by Leverrier using the least squares method.

\section{The processing of the meridian observations}\label{sec2}

Besides the unknowns $\delta\varepsilon, \delta\it n,\delta e_{0}, \delta\pi_{0}$ in the equations of condition  obtained by Leverrier, namely the corrections of the mean longitude at epoch, mean motion, eccentricity and the longitude of perihelion at epoch, we shall introduce the unknowns $\delta e'$ and $\delta \pi'$; thus
\vspace{-0.2cm}
\begin{eqnarray}
\delta e = \delta e_{0} + t \, \delta e'\,, \nonumber \\
\delta \pi = \delta \pi_{0} + t \, \delta \pi'\nonumber,
\end{eqnarray}
where $e'$ and $\pi'$  are the annual variations given by Lagrange's equations.

From the 195 equations of condition we eliminated those for which the diffe-rence between the observed apparent geocentric longitude and the calculated one is $\it{O-C}> 6"$. On the whole, we used 187 equations of condition, which have been improved by calculating the coefficients corresponding to the unknowns introduced by us.

\vspace{2cm}
The system of the 187 equations of condition is:
\vspace{-0.5cm}

\begin{eqnarray}
0.142 \delta \varepsilon   -6.9\delta n  -0.343\delta e_{0}   -0.053\delta \pi_{0}   +16.744\delta e' +2.587\delta\pi'& = &   +0.0\hspace{0.5cm} 1\nonumber \\
0.246 \delta \varepsilon  -11.9\delta n  +0.348\delta e_{0}   -0.073\delta \pi_{0}   -16.891\delta e' +3.543\delta\pi'& = &  +1.7\hspace{0.5cm} 3\nonumber \\
0.118 \delta \varepsilon   -5.7\delta n  +0.343\delta e_{0}   -0.049\delta \pi_{0}   -16.637\delta e' +2.377\delta\pi'& = &   +0.2\hspace{0.5cm} 5\nonumber \\
0.014 \delta \varepsilon   -0.7\delta n  +0.353\delta e_{0}   -0.056\delta \pi_{0}   -17.114\delta e' +2.715\delta\pi'& = &    -2.0\hspace{0.5cm}1\nonumber \\
-0.149\delta \varepsilon   +7.2\delta n  +0.547\delta e_{0}   -0.026\delta \pi_{0}   -26.450\delta e' +1.257\delta\pi'& = & +0.5\hspace{0.5cm} 6\nonumber \\
0.082 \delta \varepsilon   -4.0\delta n  +0.330\delta e_{0}   -0.048\delta \pi_{0}   -15.953\delta e' +2.321\delta\pi'& = &  -0.6\hspace{0.5cm} 5\nonumber \\
0.183 \delta \varepsilon   -8.9\delta n  +0.331\delta e_{0}   -0.063\delta \pi_{0}   -15.997\delta e' +3.238\delta\pi'& = &    -1.0\hspace{0.5cm} 2\nonumber \\
0.259 \delta \varepsilon  -12.5\delta n  +0.431\delta e_{0}   -0.062\delta \pi_{0}   -20.825\delta e' +2.996\delta\pi'& = & -1.0\hspace{0.5cm} 2\nonumber \\
0.120 \delta \varepsilon   -5.7\delta n  +0.242\delta e_{0}   -0.071\delta \pi_{0}   -11.507\delta e' +3.376\delta\pi'& = &   +0.3\hspace{0.5cm} 3\nonumber \\
0.070 \delta \varepsilon   -3.3\delta n  +0.229\delta e_{0}   -0.072\delta \pi_{0}   -10.887\delta e' +3.423\delta\pi'& = &     +1.0\hspace{0.5cm} 2\nonumber \\
-0.013\delta \varepsilon   +0.6\delta n  +0.226\delta e_{0}   -0.085\delta \pi_{0}   -10.741\delta e' +4.040\delta\pi'& = &  -1.2\hspace{0.5cm} 4\nonumber \\
-0.167\delta \varepsilon   +7.9\delta n  +0.295\delta e_{0}   -0.121\delta \pi_{0}   -14.014\delta e' +5.748\delta\pi'& = &     +1.0\hspace{0.5cm} 2\nonumber \\
-0.220\delta \varepsilon  +10.4\delta n  +0.047\delta e_{0}   +0.111\delta \pi_{0}   -2.228\delta e'  -5.261\delta\pi'& = &   +0.3\hspace{0.5cm} 4\nonumber \\
0.205 \delta \varepsilon   -9.7\delta n  +0.253\delta e_{0}   -0.079\delta \pi_{0}   -11.985\delta e' +3.742\delta\pi'& = &  -0.7\hspace{0.5cm} 4\nonumber \\
0.282 \delta \varepsilon  -13.4\delta n  +0.367\delta e_{0}   -0.083\delta \pi_{0}   -17.380\delta e' +3.930\delta\pi'& = &   -1.0\hspace{0.5cm} 4\nonumber \\
0.223 \delta \varepsilon  -10.5\delta n  +0.424\delta e_{0}   +0.057\delta \pi_{0}   -20.048\delta e' -2.695\delta\pi'& = &   +0.2\hspace{0.5cm} 4\nonumber \\
0.194 \delta \varepsilon   -9.2\delta n  +0.314\delta e_{0}   +0.075\delta \pi_{0}   -14.841\delta e' -3.545\delta\pi'& = &  +1.7\hspace{0.5cm} 3\nonumber \\
0.165 \delta \varepsilon   -7.8\delta n  +0.212\delta e_{0}   +0.083\delta \pi_{0}   -10.016\delta e' -3.921\delta\pi'& = & +0.0\hspace{0.5cm} 3\nonumber \\
0.037 \delta \varepsilon   -1.7\delta n  +0.081\delta e_{0}   +0.079\delta \pi_{0}   -3.823\delta e'  -3.729\delta\pi'& = & -1.5\hspace{0.5cm} 2\nonumber \\
-0.067\delta \varepsilon   +3.2\delta n  +0.143\delta e_{0}   +0.082\delta \pi_{0}   -6.748\delta e'  -3.869\delta\pi'& = & -4.0\hspace{0.5cm} 1\nonumber \\
0.267 \delta \varepsilon  -12.4\delta n  +0.208\delta e_{0}   -0.098\delta \pi_{0}   -9.653\delta e'    +4.548\delta \pi'& = & +1.5\hspace{0.5cm} 2\nonumber \\
0.204 \delta \varepsilon  -9.5 \delta n  +0.418\delta e_{0}   +0.051\delta \pi_{0}   -19.359\delta e'   -2.362\delta \pi'& = & +3.0\hspace{0.5cm} 1\nonumber \\
0.258 \delta \varepsilon  -11.7\delta n  +0.057\delta e_{0}   -0.103\delta \pi_{0}   -2.591\delta e'    +4.682\delta \pi'& = & +1.0\hspace{0.5cm} 3\nonumber \\
0.170 \delta \varepsilon  -7.7 \delta n  +0.407\delta e_{0}   +0.043\delta \pi_{0}   -18.456\delta e'   -1.950\delta \pi'& = & +1.0\hspace{0.5cm} 1\nonumber \\
0.064 \delta \varepsilon  -2.9 \delta n  +0.314\delta e_{0}   +0.053\delta \pi_{0}   -14.228\delta e'   -2.401\delta \pi'& = & -1.0\hspace{0.5cm} 2\nonumber \\
-0.019\delta \varepsilon  +0.9 \delta n  +0.331\delta e_{0}   +0.052\delta \pi_{0}   -14.993\delta e'   -2.355\delta \pi'& = & +1.5\hspace{0.5cm} 2\nonumber \\
0.296 \delta \varepsilon  -13.2\delta n  -0.211\delta e_{0}   -0.105\delta \pi_{0}   +9.436\delta e'    +4.697\delta \pi'& = & +1.0\hspace{0.5cm} 4\nonumber \\
0.221 \delta \varepsilon   -9.8\delta n  +0.485\delta e_{0}   -0.005\delta \pi_{0}   -21.539\delta e'   +0.222\delta \pi'& = & +3.0\hspace{0.5cm} 2\nonumber \\
0.155 \delta \varepsilon   -6.9\delta n  +0.432\delta e_{0}   +0.020\delta \pi_{0}   -19.176\delta e'   -0.888\delta \pi'& = & +4.0\hspace{0.5cm} 1\nonumber \\
-0.003\delta \varepsilon   +0.1\delta n  +0.400\delta e_{0}   +0.027\delta \pi_{0}   -17.739\delta e'   -1.197\delta \pi'& = & -2.5\hspace{0.5cm} 2\nonumber \\
0.169 \delta \varepsilon  -7.5 \delta n  +0.454\delta e_{0}   -0.007\delta \pi_{0}   -20.074\delta e'   +0.309\delta \pi'& = & +2.0\hspace{0.5cm} 1\nonumber \\
0.228 \delta \varepsilon  -10.1\delta n  +0.508\delta e_{0}   +0.021\delta \pi_{0}   -22.450\delta e'   -0.928\delta \pi'& = & -2.0\hspace{0.5cm} 2\nonumber \\
0.034 \delta \varepsilon  -1.5 \delta n  -0.319\delta e_{0}   -0.097\delta \pi_{0}   +13.948\delta e'    +4.241\delta \pi'& = & +1.5\hspace{0.5cm} 2\nonumber \\
0.305 \delta \varepsilon  -13.3\delta n  -0.139\delta e_{0}   -0.112\delta \pi_{0}   +6.052\delta e'     +4.876\delta \pi'& = & -2.0\hspace{0.5cm} 1\nonumber \\
0.217 \delta \varepsilon   -9.4\delta n +0.456 \delta e_{0}  -0.028\delta \pi_{0}   -19.803\delta e'   +1.216\delta \pi'& = & +2.0\hspace{0.5cm} 1\nonumber \\
0.087 \delta \varepsilon   -3.8\delta n  +0.390\delta e_{0}   -0.022\delta \pi_{0}   -16.878\delta e'   +0.952\delta \pi'& = & +2.0\hspace{0.5cm} 1\nonumber \\
0.199 \delta \varepsilon   -8.6\delta n  +0.043\delta e_{0}   +0.100\delta \pi_{0}   -1.855\delta e'    -4.314\delta \pi'& = & +4.0\hspace{0.5cm} 2\nonumber \\
-0.044\delta \varepsilon  +1.9 \delta n  -0.427\delta e_{0}   -0.039\delta \pi_{0}   +18.264\delta e'    +1.668\delta \pi'& = & +0.0\hspace{0.5cm} 2\nonumber \\
0.184 \delta \varepsilon   -7.8\delta n  -0.324\delta e_{0}   -0.060\delta \pi_{0}   +13.804\delta e'    +2.556\delta \pi'& = & +1.2\hspace{0.5cm} 5\nonumber \\
0.293 \delta \varepsilon  -12.5\delta n  +0.365\delta e_{0}   -0.084\delta \pi_{0}   -15.520\delta e'   +3.572\delta
\pi'& = & -2.0\hspace{0.5cm} 1\nonumber \\
0.240 \delta \varepsilon  -10.2\delta n +0.399 \delta e_{0}   -0.060\delta \pi_{0}   -16.959\delta e'   +2.550\delta \pi'& = & +0.0\hspace{0.5cm} 2\nonumber \\
0.118 \delta \varepsilon   -5.0\delta n  +0.387\delta e_{0}   -0.035\delta \pi_{0}   -16.427\delta e'   +1.486\delta \pi'& = & +1.8\hspace{0.5cm} 5\nonumber \\
0.005 \delta \varepsilon   -0.2\delta n  +0.402\delta e_{0}   -0.038\delta \pi_{0}   -17.065\delta e'   +1.613\delta \pi'& = & +2.7\hspace{0.5cm} 3\nonumber \\
-0.088\delta \varepsilon   +3.7\delta n  -0.482\delta e_{0}   -0.010\delta \pi_{0}   +20.155\delta e'    +0.175\delta \pi'& = & +3.0\hspace{0.5cm} 4\nonumber \\
0.158 \delta \varepsilon   -6.6\delta n  -0.402\delta e_{0}   -0.036\delta \pi_{0}   +16.748\delta e'    +1.500\delta \pi'& = & +3.5 \hspace{0.5cm} 2\nonumber \\
0.286 \delta \varepsilon  -11.9\delta n  -0.397\delta e_{0}   -0.077\delta \pi_{0}   +16.525\delta e'    +3.205\delta \pi'& = & +0.7\hspace{0.5cm} 4\nonumber \\
-0.209\delta \varepsilon  +8.7 \delta n  +0.443\delta e_{0}   -0.110\delta \pi_{0}  -18.373\delta e'   +4.562\delta \pi'& = & +2.0\hspace{0.5cm} 2\nonumber \\
0.225 \delta \varepsilon  -9.3 \delta n  +0.409\delta e_{0}   +0.065\delta \pi_{0}   -16.898\delta e'   -2.685\delta \pi'& = & +1.0\hspace{0.5cm} 1\nonumber \\
0.188 \delta \varepsilon  -7.8 \delta n  +0.226\delta e_{0}   +0.086\delta \pi_{0}   -9.319\delta e'    -3.546\delta \pi'& = & +2.0\hspace{0.5cm} 1\nonumber \\
0.257 \delta \varepsilon  -10.4\delta n  +0.186\delta e_{0}   -0.096\delta \pi_{0}   -7.551\delta e'    +3.897\delta \pi'& = & +0.0\hspace{0.5cm} 1\nonumber \\
0.159 \delta \varepsilon  -6.4 \delta n  +0.201\delta e_{0}   -0.079\delta \pi_{0}   -8.156\delta e'    +3.206\delta \pi'& = & +0.0\hspace{0.5cm} 1\nonumber \\
0.259 \delta \varepsilon  -10.3\delta n  +0.059\delta e_{0}   -0.104\delta \pi_{0}   -2.339\delta e'    +4.123\delta \pi'& = & +1.0\hspace{0.5cm} 1\nonumber \\
0.086 \delta \varepsilon   -3.4\delta n   +0.045\delta e_{0}  -0.086\delta \pi_{0}   -1.783\delta e'    +3.407\delta \pi'& = & +0.0\hspace{0.5cm} 1\nonumber \\
0.244 \delta \varepsilon   -9.6\delta n   +0.509\delta e_{0}  +0.015\delta \pi_{0}   -20.030\delta e'   -0.590\delta \pi'& = & +2.7\hspace{0.5cm} 3\nonumber \\
0.100 \delta \varepsilon   -3.9\delta n   -0.407\delta e_{0}  +0.023\delta \pi_{0}   +15.860\delta e'    -0.896\delta \pi'& = & +1.5\hspace{0.5cm} 2\nonumber \\
0.240 \delta \varepsilon   -9.3\delta n   -0.511\delta e_{0}  +0.011\delta \pi_{0}   +19.811\delta e'    -0.426\delta \pi'& = & +4.0\hspace{0.5cm} 2\nonumber \\
0.254 \delta \varepsilon   -9.8\delta n   -0.524\delta e_{0}  +0.001\delta \pi_{0}   +20.311\delta e'    -0.039\delta \pi'& = & +0.0\hspace{0.5cm} 2\nonumber \\
0.002 \delta \varepsilon   -0.1\delta n   +0.360\delta e_{0}  +0.043\delta \pi_{0}   -13.795\delta e'   -1.648\delta \pi'& = & +0.2\hspace{0.5cm} 4\nonumber \\
-0.078\delta \varepsilon   +3.0\delta n   +0.411\delta e_{0}  +0.041 \delta \pi_{0}  -15.744\delta e'   -1.571\delta \pi'& = & +1.3\hspace{0.5cm} 3\nonumber \\
0.231 \delta \varepsilon   -8.6\delta n   +0.485\delta e_{0}  -0.020\delta \pi_{0}   -18.155\delta e'   +0.749\delta
\pi'& = & +3.0\hspace{0.5cm} 1\nonumber \\
0.176 \delta \varepsilon   -6.6\delta n   +0.452\delta e_{0}  +0.002\delta \pi_{0}   -16.912\delta e'   -0.075\delta \pi'& = & -1.0\hspace{0.5cm} 1\nonumber \\
-0.125\delta \varepsilon   +4.7\delta n   +0.517\delta e_{0}  +0.007\delta \pi_{0}   -19.311\delta e'   -0.261\delta \pi'& = & -4.5\hspace{0.5cm} 2\nonumber \\
0.251 \delta \varepsilon   -9.1\delta n   +0.436\delta e_{0}  -0.053\delta \pi_{0}   -15.901\delta e'   +1.933\delta \pi'& = & +1.0\hspace{0.5cm} 1\nonumber \\
-0.001\delta \varepsilon   +0.0\delta n   +0.427\delta e_{0}  -0.019\delta \pi_{0}   -15.550\delta e'   +0.692\delta \pi'& = & +0.3\hspace{0.5cm} 3\nonumber \\
-0.073\delta \varepsilon   +2.7\delta n   +0.471\delta e_{0}  -0.022\delta \pi_{0}   -17.149\delta e'   +0.801\delta \pi'& = & +1.0\hspace{0.5cm} 1\nonumber \\
0.193 \delta \varepsilon   -7.0\delta n   +0.420\delta e_{0}  -0.041\delta \pi_{0}   -15.237\delta e'   +1.487\delta \pi'& = & +4.0\hspace{0.5cm} 1\nonumber \\
0.188 \delta \varepsilon   -6.1\delta n   +0.031\delta e_{0}  -0.091\delta \pi_{0}   -1.012\delta e'    +2.971\delta \pi'& = & +0.0\hspace{0.5cm} 1\nonumber \\
-0.126\delta \varepsilon   +4.0\delta n   +0.490\delta e_{0}  +0.028\delta \pi_{0}   -15.348\delta e'   -0.877\delta \pi'& = & +6.0\hspace{0.5cm} 1\nonumber \\
0.093 \delta \varepsilon   -2.9\delta n   -0.227\delta e_{0}  -0.071\delta \pi_{0}   +6.976\delta e'     +2.182\delta \pi'& = & +2.3\hspace{0.5cm} 3\nonumber \\
0.140 \delta \varepsilon   -4.2\delta n   -0.292\delta e_{0}   -0.064\delta \pi_{0}   +8.696\delta e'     +1.906\delta \pi'& = & +3.0\hspace{0.5cm} 1\nonumber \\
0.160 \delta \varepsilon   -4.7\delta n   -0.306\delta e_{0}   -0.063\delta \pi_{0}   +9.060\delta e'     +1.865\delta \pi'& = & +2.0\hspace{0.5cm} 1\nonumber \\
0.267 \delta \varepsilon   -7.9\delta n   +0.401\delta e_{0}   -0.069\delta \pi_{0}   -11.832\delta e'   +2.036\delta \pi'& = & -2.0\hspace{0.5cm} 2\nonumber \\
0.112 \delta \varepsilon   -3.3\delta n   +0.396\delta e_{0}   -0.030\delta \pi_{0}   -11.671\delta e'   +0.884\delta \pi'& = & +0.5\hspace{0.5cm} 2\nonumber \\
0.064 \delta \varepsilon   -1.8\delta n   +0.323\delta e_{0}   -0.049\delta \pi_{0}   -9.159\delta e'    +1.389\delta \pi'& = & +4.0\hspace{0.5cm} 3\nonumber \\
0.058 \delta \varepsilon   -1.6\delta n   +0.238\delta e_{0}   +0.067\delta \pi_{0}   -6.677\delta e'    -1.880\delta \pi'& = & -2.0\hspace{0.5cm} 1\nonumber \\
0.142 \delta \varepsilon   -4.0\delta n   -0.412\delta e_{0}   -0.028\delta \pi_{0}   +11.484\delta e'    +0.780\delta \pi'& = & -1.0\hspace{0.5cm} 1\nonumber \\
-0.092\delta \varepsilon   +2.6\delta n   -0.468\delta e_{0}   +0.011\delta \pi_{0}   +13.036\delta e'    -0.306\delta \pi'& = & +0.0 \hspace{0.5cm} 1\nonumber \\
0.215 \delta \varepsilon   -5.9\delta n   +0.223\delta e_{0}   -0.084\delta \pi_{0}   -6.151\delta e'    +2.317\delta \pi'& = & +0.0\hspace{0.5cm} 1\nonumber \\
0.123 \delta \varepsilon   -3.4\delta n   +0.209\delta e_{0}   -0.075\delta \pi_{0}   -5.761\delta e'    +2.067\delta \pi'& = & -2.3\hspace{0.5cm} 3\nonumber \\
-0.014\delta \varepsilon   +0.4\delta n   +0.183\delta e_{0}   -0.090\delta \pi_{0}   -5.041\delta e'    +2.479\delta
\pi'& = & -1.0\hspace{0.5cm} 2\nonumber \\
0.232 \delta \varepsilon   -6.2\delta n   -0.496\delta e_{0}   -0.010\delta \pi_{0}   +13.255\delta e'    +0.267\delta \pi'& = & +0.5\hspace{0.5cm} 2\nonumber \\
0.019 \delta \varepsilon   -0.5\delta n   +0.031\delta e_{0}   -0.092\delta \pi_{0}   -0.825\delta e'    +2.447\delta \pi'& = & -1.0\hspace{0.5cm} 2\nonumber \\
0.254 \delta \varepsilon   -6.7\delta n   +0.525\delta e_{0}   +0.009\delta \pi_{0}   -13.834\delta e'   -0.237\delta \pi'& = & +0.5\hspace{0.5cm} 2\nonumber \\
0.238 \delta \varepsilon   -6.3\delta n   +0.502\delta e_{0}   +0.023\delta \pi_{0}   -13.226\delta e'   -0.606\delta \pi'& = & -0.5\hspace{0.5cm} 2\nonumber \\
0.198 \delta \varepsilon   -5.2\delta n   +0.418\delta e_{0}   +0.048\delta \pi_{0}   -11.003\delta e'   -1.263\delta \pi'& = & +0.7\hspace{0.5cm} 3\nonumber \\
0.166 \delta \varepsilon   -4.4\delta n   +0.349\delta e_{0}   +0.059\delta \pi_{0}   -9.182\delta e'    -1.552\delta \pi'& = & +0.3\hspace{0.5cm} 3\nonumber \\
0.194 \delta \varepsilon   -5.0\delta n   -0.021\delta e_{0}   -0.093\delta \pi_{0}   +0.539\delta e'     +2.387\delta \pi'& = & -0.5\hspace{0.5cm} 2\nonumber \\
0.259 \delta \varepsilon   -6.6\delta n   +0.023\delta e_{0}   -0.104\delta \pi_{0}   -0.586\delta e'    +2.649\delta \pi'& = & +0.0\hspace{0.5cm} 1\nonumber \\
0.326 \delta \varepsilon   -8.3\delta n   +0.188\delta e_{0}   -0.113\delta \pi_{0}   -4.785\delta e'    +2.876\delta \pi'& = & -1.0\hspace{0.5cm} 2\nonumber \\
0.258 \delta \varepsilon   -6.6\delta n   +0.519\delta e_{0}   -0.013\delta \pi_{0}   -13.174\delta e'   +0.330\delta \pi'& = & +4.0\hspace{0.5cm} 2\nonumber \\
0.131 \delta \varepsilon   -3.3\delta n   +0.374\delta e_{0}   +0.044\delta \pi_{0}   -9.479\delta e'    -1.115\delta \pi'& = & +2.5\hspace{0.5cm} 2\nonumber \\
0.314 \delta \varepsilon   -7.8\delta n   -0.324\delta e_{0}   -0.098\delta \pi_{0}   +8.016\delta e'     +2.425\delta \pi'& = & +3.7\hspace{0.5cm} 3\nonumber \\
0.277 \delta \varepsilon   -6.8\delta n   -0.213\delta e_{0}   -0.099\delta \pi_{0}   +5.268\delta e'     +2.448\delta \pi'& = & +2.3\hspace{0.5cm} 3\nonumber \\
-0.050\delta \varepsilon   +1.2\delta n   +0.447\delta e_{0}   +0.014\delta \pi_{0}   -10.886\delta e'   -0.341\delta \pi'& = & -2.0\hspace{0.5cm} 2\nonumber \\
0.317 \delta \varepsilon   -7.4\delta n   +0.383\delta e_{0}   -0.089\delta \pi_{0}   -8.999\delta e'    +2.091\delta \pi'& = & +2.7\hspace{0.5cm} 3\nonumber \\
0.177 \delta \varepsilon   -4.1\delta n   +0.434\delta e_{0}   -0.023\delta \pi_{0}   -10.179\delta e'   +0.539\delta \pi'& = & +1.0\hspace{0.5cm} 1\nonumber \\
0.054 \delta \varepsilon   -1.3\delta n   +0.387\delta e_{0}   -0.022\delta \pi_{0}   -9.012\delta e'    +0.512\delta \pi'& = & +4.0\hspace{0.5cm} 1\nonumber \\
0.187 \delta \varepsilon   -4.3\delta n   +0.422\delta e_{0}   -0.038\delta \pi_{0}   -9.820\delta e'    +0.884\delta \pi'& = & +0.5\hspace{0.5cm} 2\nonumber \\
-0.086\delta \varepsilon   +2.0\delta n   -0.401\delta e_{0}   -0.073\delta \pi_{0}   +9.091\delta e'     +1.655\delta \pi'& = & +0.0\hspace{0.5cm} 1\nonumber \\
0.329 \delta \varepsilon   -7.4\delta n   -0.232\delta e_{0}   -0.111\delta \pi_{0}   +5.239\delta e'     +2.507\delta \pi'& = & -5.0\hspace{0.5cm} 1\nonumber \\
0.107 \delta \varepsilon   -2.4\delta n   +0.367\delta e_{0}   -0.042\delta \pi_{0}   -8.252\delta e'    +0.944\delta
\pi'& = & +4.0\hspace{0.5cm} 1\nonumber \\
-0.058\delta \varepsilon   +1.3\delta n   +0.414\delta e_{0}   -0.056\delta \pi_{0}   -9.926\delta e'    +1.257\delta \pi'& = & -3.0\hspace{0.5cm} 1\nonumber \\
0.197 \delta \varepsilon   -4.4\delta n   +0.185\delta e_{0}   +0.092\delta \pi_{0}   -4.108\delta e'    -2.043\delta \pi'& = & +2.0\hspace{0.5cm} 1\nonumber \\
0.048 \delta \varepsilon   -1.0\delta n   +0.042\delta e_{0}   +0.081\delta \pi_{0}   -0.890\delta e'    -1.717\delta \pi'& = & -4.0\hspace{0.5cm} 3\nonumber \\
0.289 \delta \varepsilon   -4.0\delta n   -0.497\delta e_{0}   -0.051 \delta \pi_{0}  +6.807\delta e'     +0.698\delta \pi'& = & -3.3\hspace{0.5cm} 1\nonumber \\
0.273 \delta \varepsilon   -3.7\delta n   +0.085\delta e_{0}   -0.106\delta \pi_{0}   -1.159\delta e'    +1.445\delta \pi'& = & -0.2\hspace{0.5cm} 4\nonumber \\
0.183 \delta \varepsilon   -2.5\delta n   +0.105\delta e_{0}   -0.089\delta \pi_{0}   -1.430\delta e'    +1.212\delta \pi'& = & -5.2\hspace{0.5cm} 4\nonumber \\
-0.005\delta \varepsilon   +0.1\delta n   +0.245\delta e_{0}   -0.078\delta \pi_{0}   -3.295\delta e'    +1.049\delta \pi'& = & -1.0\hspace{0.5cm} 1\nonumber \\
0.210 \delta \varepsilon   -2.8\delta n   +0.134\delta e_{0}   -0.092\delta \pi_{0}   -1.799\delta e'    +1.235\delta \pi'& = & -3.6\hspace{0.5cm} 1\nonumber \\
0.230 \delta \varepsilon   -3.1\delta n   +0.485\delta e_{0}   +0.031\delta \pi_{0}   -6.480\delta e'    -0.414\delta \pi'& = & -0.6\hspace{0.5cm} 2\nonumber \\
0.033 \delta \varepsilon   -0.4\delta n   +0.218\delta e_{0}   +0.070\delta \pi_{0}   -2.891\delta e'    -0.928\delta \pi'& = & +0.3\hspace{0.5cm} 1\nonumber \\
0.148 \delta \varepsilon   -1.9\delta n   +0.409\delta e_{0}   +0.038\delta \pi_{0}   -5.369\delta e'    -0.499\delta \pi'& = & +2.8\hspace{0.5cm} 2\nonumber \\
0.106 \delta \varepsilon   -1.4\delta n   -0.342\delta e_{0}   +0.051\delta \pi_{0}   +4.381\delta e'     -0.653\delta \pi'& = & +1.2\hspace{0.5cm} 1\nonumber \\
0.260 \delta \varepsilon   -3.3\delta n   -0.526\delta e_{0}   -0.009\delta \pi_{0}   +6.705\delta e'     +0.115\delta \pi'& = & -1.1\hspace{0.5cm} 2\nonumber \\
-0.141\delta \varepsilon   +1.8\delta n   -0.212\delta e_{0}   -0.119\delta \pi_{0}   +2.676\delta e'     +1.502\delta \pi'& = & +0.9\hspace{0.5cm} 1\nonumber \\
0.040 \delta \varepsilon   -0.5\delta n   +0.084\delta e_{0}   -0.086\delta \pi_{0}   -1.049\delta e'    +1.074\delta \pi'& = & +2.2\hspace{0.5cm} 2\nonumber \\
0.281 \delta \varepsilon   -3.5\delta n   +0.530\delta e_{0}   -0.030\delta \pi_{0}   -6.571\delta e'    +0.372\delta \pi'& = & +0.7\hspace{0.5cm} 1\nonumber \\
0.219 \delta \varepsilon   -2.7\delta n   +0.485\delta e_{0}   +0.017\delta \pi_{0}   -6.002\delta e'    -0.210\delta \pi'& = & +0.5\hspace{0.5cm} 2\nonumber \\
0.164 \delta \varepsilon   -2.0\delta n   +0.404\delta e_{0}   +0.041\delta \pi_{0}   -4.989\delta e'    -0.506\delta \pi'& = & +0.4\hspace{0.5cm} 4\nonumber \\
0.015 \delta \varepsilon   -0.2\delta n   +0.329\delta e_{0}   +0.050\delta \pi_{0}   -4.049\delta e'    -0.615\delta \pi'& = & +0.3\hspace{0.5cm} 2\nonumber \\
-0.257\delta \varepsilon   +3.1\delta n   +0.583\delta e_{0}   +0.053\delta \pi_{0}   -7.155\delta e'    -0.650\delta
\pi'& = & -5.4\hspace{0.5cm} 2\nonumber \\
-0.425\delta \varepsilon   +5.2\delta n   +0.616\delta e_{0}   +0.122\delta \pi_{0}   -7.527\delta e'    -1.491\delta \pi'& = & +0.6\hspace{0.5cm} 1\nonumber \\
0.211 \delta \varepsilon   -2.6\delta n   +0.465\delta e_{0}   +0.039\delta \pi_{0}   -5.652\delta e'    -0.474\delta \pi'& = & -0.2\hspace{0.5cm} 2\nonumber \\
0.175 \delta \varepsilon   -2.1\delta n   -0.245\delta e_{0}   +0.080\delta \pi_{0}   +2.950\delta e'    -0.963\delta \pi'& = & -1.4\hspace{0.5cm} 1\nonumber \\
0.087 \delta \varepsilon   -1.0\delta n   -0.373\delta e_{0}   +0.038\delta \pi_{0}   +4.476\delta e'     -0.456\delta \pi'& = & +0.0\hspace{0.5cm} 2\nonumber \\
-0.076\delta \varepsilon   +0.9\delta n   -0.269\delta e_{0}   +0.071\delta \pi_{0}   +3.200\delta e'     -0.845\delta \pi'& = & +4.7\hspace{0.5cm} 1\nonumber \\
0.170 \delta \varepsilon   -2.0\delta n   -0.336\delta e_{0}   +0.064\delta \pi_{0}   +3.976\delta e'     -0.757\delta \pi'& = & +3.6\hspace{0.5cm} 1\nonumber \\
0.288 \delta \varepsilon   -3.4\delta n   -0.214\delta e_{0}   -0.103\delta \pi_{0}   +2.509\delta e'     +1.208\delta \pi'& = & -1.9\hspace{0.5cm} 1\nonumber \\
-0.313\delta \varepsilon   +3.6\delta n   -0.034\delta e_{0}   -0.184\delta \pi_{0}   +0.393\delta e'     +2.131\delta \pi'& = & -3.8\hspace{0.5cm} 1\nonumber \\
-0.132\delta \varepsilon   +1.5\delta n   +0.005\delta e_{0}   -0.125\delta \pi_{0}   -0.058\delta e'    +1.445\delta \pi'& = & +3.9\hspace{0.5cm} 1\nonumber \\
0.132 \delta \varepsilon   -1.5\delta n   +0.417\delta e_{0}   +0.020\delta \pi_{0}   -4.748\delta e'    -0.228\delta \pi'& = & -0.3\hspace{0.5cm} 3\nonumber \\
0.104 \delta \varepsilon   -1.2\delta n   +0.405\delta e_{0}   +0.023\delta \pi_{0}   -4.608\delta e'    -0.262\delta \pi'& = & -1.9\hspace{0.5cm} 3\nonumber \\
0.020 \delta \varepsilon   -0.2\delta n   +0.397\delta e_{0}   +0.025\delta \pi_{0}   -4.509\delta e'    -0.284\delta \pi'& = & -1.0\hspace{0.5cm} 1\nonumber \\
-0.112\delta \varepsilon   +1.3\delta n   +0.492\delta e_{0}   +0.018\delta \pi_{0}   -5.579\delta e'    -0.204\delta \pi'& = & -1.7\hspace{0.5cm} 2\nonumber \\
-0.211\delta \varepsilon   +2.4\delta n   +0.604\delta e_{0}   +0.015\delta \pi_{0}   -6.841\delta e'    -0.170\delta \pi'& = & -0.6\hspace{0.5cm} 2\nonumber \\
-0.037\delta \varepsilon   +0.4\delta n   +0.410\delta e_{0}   +0.017\delta \pi_{0}   -4.609\delta e'    -0.191\delta \pi'& = & -0.3\hspace{0.5cm} 2\nonumber \\
0.180 \delta \varepsilon   -2.0\delta n   +0.463\delta e_{0}   -0.010\delta \pi_{0}   -5.194\delta e'   +0.112\delta \pi'& = & -0.1\hspace{0.5cm} 2\nonumber \\
0.233 \delta \varepsilon   -2.6\delta n   +0.510\delta e_{0}   +0.020\delta \pi_{0}   -5.710\delta e'    -0.224\delta \pi'& = & -1.3\hspace{0.5cm} 2\nonumber \\
0.064 \delta \varepsilon   -0.7\delta n   -0.297\delta e_{0}   +0.058\delta \pi_{0}   +3.280\delta e'     -0.640\delta \pi'& = & -2.1\hspace{0.5cm} 2\nonumber \\
-0.183\delta \varepsilon   +2.0\delta n   -0.247\delta e_{0}   +0.094\delta \pi_{0}   +2.706\delta e'     -1.030\delta \pi'& = & +5.1\hspace{0.5cm} 2\nonumber \\
0.239 \delta \varepsilon   -2.6\delta n   -0.477\delta e_{0}   +0.044\delta \pi_{0}   +5.171\delta e'     -0.477\delta
\pi'& = & +4.2\hspace{0.5cm} 1\nonumber \\
0.257 \delta \varepsilon   -2.8\delta n   -0.522\delta e_{0}   +0.026\delta \pi_{0}   +5.654\delta e'     -0.282\delta \pi'& = & -0.1\hspace{0.5cm} 5\nonumber \\
-0.077\delta \varepsilon   +0.8\delta n   -0.455\delta e_{0}   -0.069\delta \pi_{0}   +4.882\delta e'     +0.740\delta \pi'& = & -0.6\hspace{0.5cm} 2\nonumber \\
0.040 \delta \varepsilon   -0.4\delta n   -0.196\delta e_{0}   -0.081\delta \pi_{0}   +2.094\delta e'     +0.865\delta \pi'& = & -0.2\hspace{0.5cm} 3\nonumber \\
0.182 \delta \varepsilon   -1.9\delta n   -0.233\delta e_{0}   -0.077\delta \pi_{0}   +2.462\delta e'     +0.814\delta \pi'& = & +2.9\hspace{0.5cm} 1\nonumber \\
0.265 \delta \varepsilon   -2.8\delta n   -0.213\delta e_{0}   -0.097\delta \pi_{0}   +2.247\delta e'     +1.023\delta \pi'& = & -1.4\hspace{0.5cm} 3\nonumber \\
0.326 \delta \varepsilon   -3.5\delta n   -0.106\delta e_{0}   -0.118\delta \pi_{0}   +1.117\delta e'     +1.243\delta \pi'& = & -4.8\hspace{0.5cm} 2\nonumber \\
0.216 \delta \varepsilon   -2.3\delta n +0.451\delta e_{0}   -0.031\delta \pi_{0}   -4.716\delta e'    +0.324\delta \pi'& = & -1.8\hspace{0.5cm} 2\nonumber \\
0.045 \delta \varepsilon   -0.5\delta n +0.417\delta e_{0}   -0.006\delta \pi_{0}   -4.343\delta e'    +0.062\delta \pi'& = & -1.5\hspace{0.5cm} 1\nonumber \\
-0.077\delta \varepsilon   +0.2\delta n +0.443\delta e_{0}   -0.010\delta \pi_{0}   -4.608\delta e'    +0.104\delta \pi'& = & -2.2\hspace{0.5cm} 2\nonumber \\
-0.147\delta \varepsilon   +1.5\delta n +0.550\delta e_{0}   -0.022\delta \pi_{0}   -5.711\delta e'    +0.228\delta \pi'& = & -4.7\hspace{0.5cm} 1\nonumber \\
-0.092\delta \varepsilon   +0.9\delta n +0.467\delta e_{0}   +0.008\delta \pi_{0}   -4.801\delta e'    -0.082\delta \pi'& = & -0.6\hspace{0.5cm} 1\nonumber \\
-0.201\delta \varepsilon   +2.0\delta n -0.065\delta e_{0}   +0.108\delta \pi_{0}   +0.654\delta e'     -1.087\delta \pi'& = & +3.2\hspace{0.5cm} 1\nonumber \\
-0.254\delta \varepsilon   +2.5\delta n -0.536\delta e_{0}   -0.109\delta \pi_{0}   +5.190\delta e'     +1.055\delta \pi'& = & -2.4\hspace{0.5cm} 1\nonumber \\
-0.134\delta \varepsilon   +1.3\delta n -0.414\delta e_{0}   -0.085\delta \pi_{0}   +4.002\delta e'     +0.822\delta \pi'& = & -1.0\hspace{0.5cm} 3\nonumber \\
0.336 \delta \varepsilon   -3.2\delta n -0.178\delta e_{0}   -0.125 \delta \pi_{0}  +1.707\delta e'     +1.199\delta \pi'& = & -5.5\hspace{0.5cm} 2\nonumber \\
0.327 \delta \varepsilon   -3.1\delta n +0.294\delta e_{0}   -0.105\delta \pi_{0}   -2.806\delta e'    +1.002\delta \pi'& = & -5.3\hspace{0.5cm} 2\nonumber \\
0.288 \delta \varepsilon   -2.7\delta n +0.358\delta e_{0}   -0.084\delta \pi_{0}   -3.410\delta e'    +0.800\delta \pi'& = & -5.6\hspace{0.5cm} 3\nonumber \\
0.020 \delta \varepsilon   -0.2\delta n +0.386\delta e_{0}   -0.041\delta \pi_{0}   -3.651\delta e'    +0.388\delta \pi'& = & -2.1\hspace{0.5cm} 2\nonumber \\
-0.036\delta \varepsilon   +0.3\delta n +0.425\delta e_{0}   -0.023 \delta \pi_{0}  -3.966\delta e'    +0.215\delta
\pi'& = & +5.1\hspace{0.5cm} 1\nonumber \\
0.152 \delta \varepsilon   -1.4\delta n +0.350\delta e_{0}   -0.053 \delta \pi_{0}  -3.261\delta e'   +0.494\delta \pi'& = & +1.7\hspace{0.5cm} 1\nonumber \\
0.225 \delta \varepsilon   -2.1\delta n +0.358\delta e_{0}   +0.076 \delta \pi_{0}  -3.302\delta e'   -0.701\delta \pi'& = & -1.5\hspace{0.5cm} 2\nonumber \\
0.211 \delta \varepsilon   -2.0\delta n +0.267\delta e_{0}   +0.087 \delta \pi_{0}  -2.460\delta e'   -0.746\delta \pi'& = & -1.7\hspace{0.5cm} 1\nonumber \\
0.113 \delta \varepsilon   -1.0\delta n +0.176\delta e_{0}   +0.069\delta \pi_{0}   -1.616\delta e'  -0.634\delta \pi'& = & -5.3\hspace{0.5cm} 1\nonumber \\
0.152 \delta \varepsilon   -1.4\delta n -0.031\delta e_{0}   +0.093\delta \pi_{0}    +0.284\delta e'  -0.853\delta \pi'& = & +0.6\hspace{0.5cm} 1\nonumber \\
-0.321\delta \varepsilon   +2.8\delta n -0.778\delta e_{0}   +0.014\delta \pi_{0}    +6.852\delta e'  -0.123\delta \pi'& = & -2.9\hspace{0.5cm} 1\nonumber \\
0.139 \delta \varepsilon   -1.2\delta n -0.407\delta e_{0}   -0.030\delta \pi_{0}    +3.529\delta e'  +0.260\delta \pi'& = & +0.0\hspace{0.5cm} 4\nonumber \\
0.258 \delta \varepsilon   -2.2\delta n -0.416\delta e_{0}   -0.062\delta \pi_{0}    +3.594\delta e'  +0.536\delta \pi'& = & -1.8\hspace{0.5cm} 1\nonumber \\
0.294 \delta \varepsilon   -2.5\delta n -0.393\delta e_{0}   -0.077\delta \pi_{0}    +3.390\delta e'  +0.664\delta \pi'& = & -5.1\hspace{0.5cm} 2\nonumber \\
0.163 \delta \varepsilon  -1.4 \delta n +0.297\delta e_{0}   -0.065\delta \pi_{0}   -2.536\delta e'  +0.555\delta \pi'& = & -1.5\hspace{0.5cm} 1\nonumber \\
0.004 \delta \varepsilon  +0.0 \delta n +0.290\delta e_{0}   -0.072\delta \pi_{0}   -2.468\delta e'  +0.613\delta \pi'& = & +2.2\hspace{0.5cm} 1\nonumber \\
0.120 \delta \varepsilon   -1.0\delta n +0.282\delta e_{0}   -0.062\delta \pi_{0}   -2.359\delta e'  +0.518\delta \pi'& = & +0.6\hspace{0.5cm} 2\nonumber \\
0.336 \delta \varepsilon   -2.8\delta n +0.470\delta e_{0}   -0.048\delta \pi_{0}   -3.890\delta e'  +0.397\delta \pi'& = & +2.8\hspace{0.5cm} 1\nonumber \\
-0.019\delta \varepsilon   +0.1\delta n +0.219\delta e_{0}   +0.069\delta \pi_{0}   -1.769\delta e'  -0.557\delta \pi'& = & -4.5\hspace{0.5cm} 1\nonumber \\
0.134 \delta \varepsilon   -1.1\delta n +0.280\delta e_{0}   +0.069\delta \pi_{0}   -2.465\delta e'  -0.556\delta \pi'& = & +1.1\hspace{0.5cm} 1\nonumber \\
0.200 \delta \varepsilon  -1.6 \delta n -0.470\delta e_{0}  -0.023\delta \pi_{0}   +3.706\delta e'   +0.181\delta \pi'& = & -1.1\hspace{0.5cm} 2\nonumber \\
0.014 \delta \varepsilon  -0.1 \delta n -0.407\delta e_{0}  -0.004\delta \pi_{0}   +3.205\delta e'   +0.031\delta \pi'& = & -1.6\hspace{0.5cm} 1\nonumber \\
0.121 \delta \varepsilon  -0.9 \delta n -0.427\delta e_{0}  +0.002\delta \pi_{0}    +3.299\delta e'  -0.015\delta \pi'& = & +4.4\hspace{0.5cm} 4\nonumber \\
0.202 \delta \varepsilon   -1.6\delta n -0.464\delta e_{0}  -0.015\delta \pi_{0}   +3.573\delta e'   +0.155\delta \pi'& = & +1.8\hspace{0.5cm} 4\nonumber \\
0.236 \delta \varepsilon  -1.8 \delta n -0.474\delta e_{0}  -0.030\delta \pi_{0}   + 3.645\delta e'  +0.231\delta \pi'& = & -0.1\hspace{0.5cm} 2\nonumber \\
0.311 \delta \varepsilon   -2.4\delta n -0.439\delta e_{0}  -0.077\delta \pi_{0}   +3.365\delta e'   +0.590\delta \pi'& = & +0.3\hspace{0.5cm} 1\nonumber \\
0.328 \delta \varepsilon   -2.5\delta n +0.038\delta e_{0}  -0.119\delta \pi_{0}  -0.290\delta e'    +0.907\delta \pi'& = & -2.5\hspace{0.5cm} 3\nonumber \\
0.117 \delta \varepsilon   -0.9\delta n +0.173\delta e_{0}  -0.079\delta \pi_{0}  -1.311\delta e'    +0.599\delta \pi'& = & -5.9\hspace{0.5cm} 1\nonumber \\
0.057 \delta \varepsilon   -0.4\delta n +0.151\delta e_{0}  -0.081\delta \pi_{0}  -1.143\delta e'    +0.613\delta \pi'& = & -4.3\hspace{0.5cm} 2\nonumber \\
-0.046\delta \varepsilon   +0.3\delta n +0.134\delta e_{0}  -0.099\delta \pi_{0}  -1.012\delta e'    +0.748\delta \pi'& = & -3.3\hspace{0.5cm} 2\nonumber \\
0.280 \delta \varepsilon   -2.1\delta n +0.288\delta e_{0}  -0.093\delta \pi_{0}  -2.127\delta e'    +0.687\delta \pi'& = & -2.9\hspace{0.5cm} 3\nonumber \\
0.309 \delta \varepsilon   -2.3\delta n +0.408\delta e_{0}  -0.084\delta \pi_{0}  -3.008\delta e'    +0.619\delta \pi'& = & -4.0\hspace{0.5cm} 4,\nonumber
\end{eqnarray}
where the last column represents the weight for each equation of condition.

The system of the 187 equations of condition is solved by the least squares method.

The general solution is:
\vspace{-0.5cm}

\begin{eqnarray}\label{1.5}
\hspace{0.5cm}\delta \varepsilon & = & -2".65 \pm 1".35,\nonumber \\
\delta n \, & = & -0".14 \pm 0".05, \nonumber\\
\delta e_{0} & = & -2".07 \pm 0".66, \nonumber \\
\delta \pi_{0} & = & 18".88 \pm 3".93, \nonumber \\
\delta e' & = & -0".044 \pm 0".022, \nonumber \\
\delta \pi' &  = & 0".428\pm 0".126.
\end{eqnarray}

In the general solution we find the values $\delta e_{0} = -2".07$ and $\delta \pi_{0} = 18".88$. These values are of the same order as the ones corresponding to the Leverrier's relation for $\delta e$ and $\delta \pi$ based on transits, namely $\delta \pi + 2.72\,\delta \hspace{-0.1cm}\ e= 10".3$ \cite{5}.

Also, in the general case we obtain $\delta e' = -0".044$. If we replace this value in the relation corresponding to the transits (\ref{1.4}), namely $\delta\pi' + 2.72\,\delta e'= 0".383$, we get an advance of $\delta\pi'= 50"/century $.

Furthermore, in the general solution (\ref{1.5}) we obtain for the Mercury's perihelion advance $\delta\pi'= 42".8/century $.

\vspace{0.1cm}
On the other hand, if we take $\delta e' = 0$, like Leverrier, who considered $\delta e'$ negligible in the relation (\ref{1.4}), then solution becomes:

\vspace{-0.5cm}
\begin{eqnarray}\label{1.7}
\delta \varepsilon & = & -2".90 \pm 1".35,\nonumber \\
\delta n\, & = & -0".16 \pm 0".05,\nonumber \\
\delta e_{0} & = & -0".95 \pm 0".34,\nonumber \\
\delta \pi_{0} & = & 18".72 \pm 3".94,\nonumber \\
\delta e' & = & 0,\nonumber\\
\delta \pi' & = & 0".428\pm 0".127.
\end{eqnarray}

We observe in the case $\delta e' = 0$ that the perihelion advance is also $\delta\pi'= 42".8/century$.

\section{Concluding remarks}\label{sec3}

The value of $\delta e'= -0".044/year$, obtained in the present paper in the general case (\ref{1.5}), is in good agreement with the Leverrier's suggestion regarding $\delta e'$, namely that is negligible. More over, this value of  $\delta e'= -0".044$ is lower in absolute value than Leverrier's estimation of $\delta e'= -0".0806$.

The result obtained by us for Mercury's perihelion advance $\delta\pi'= 42".8/cen-tury $, by processing the meridian observations used by Leverrier, is very close both to the values obtained by Newcomb and Clemence and to the value calculated within the theory of general relativity $\delta\pi'= 42".98/century $ \cite{7}. This value obtained in a relativistic frame is also confirmed by radar data processing, where the value $\delta\pi'= 42".94/century $ is obtained \cite{BG},\cite{8}.

\vspace{0.8cm}
\textbf{Acknowledgments}. I should like to express my gratitude to my supervisor Prof. Dr. Ieronim Mihaila, from University of Bucharest, and to my colleagues Dr. Dumitru Pricopi and Dr. Doru Suran for useful conversations and remarks.

\vspace{0.5cm}

{\it Diana Rodica Constantin}\\
University of Bucharest, Faculty of Mathematics and Computer Science, \\
14 Academiei Str., 010014 Bucharest, Romania \\
E-mail: {\tt ghe12constantin@yahoo.com}

\end{document}